\documentclass[Draft]{jhep3}

\usepackage{mathrsfs}
\usepackage{amsmath}
\usepackage{graphicx}
\usepackage{subfigure}
\usepackage{epstopdf}

\title{QCD and the Hagedorn spectrum}

\author{Thomas D. Cohen\\Department of Physics\\ University of Maryland\\  College Park, MD 20742-4111       \\Email: \email{cohen@physics.umd.edu}}

\abstract{It is shown that large $N_c$ QCD  must have a  Hagedorn spectrum ({\it i.e.,} a spectrum of hadron which grows exponentially with the hadrons' mass) provided that certain technical assumptions concerning the applicability of perturbation theory to a certain class of correlation functions apply.  The basic argument exploits the interplay of confinement and asymptotic freedom.}

\begin{document}
\section{Introduction}
More than four decades ago, Hagedorn \cite{Hag} proposed that the number of hadrons with mass less than $m$ grew exponentially with $m$. Neglecting hadron-hadron interactions this implies that the energy density diverges at a finite temperature, yielding a maximum temperature for hadronic matter, $T_H$.  A Hagedorn spectrum was found to arise automatically in string theories \cite{Pol} (which were originally formulated as a theory of the strong interaction).  Thus, the idea of the Hagedorn spectrum fits naturally with our understanding of QCD:  QCD is confining and as such one might expect it to have stringy dynamics for high-lying states. Moreover, at high temperatures QCD is in a quark-gluon phase rather than a hadronic phase.

Despite empirical evidence that the number of hadrons up to  $\sim 2$ GeV does grow rapidly with the mass in a  manner qualitatively consistent with an exponential  growth\cite{spec}, the picture described above is clearly imprecise.  Hadron masses are not strictly well defined: hadrons have widths due to strong decays.  Moreover, unlike in pure Yang-Mills theory, there is no order parameter for deconfinement in QCD;  as such, the distinction between a hadronic phase and a deconfined phase is not sharp.  However, the notion of a Hagedorn spectrum becomes precise if applied to the large $N_c$ limit of QCD\cite{largeN1,largeN2} rather than QCD itself.  In the large $N_c$ limit meson and glueball widths go to zero  making the hadron masses well defined\cite{largeN1,largeN2}.   Moreover, at large $N_c$, QCD has a first order phase transition to a deconfining phase\cite{largeNthermo}.

The conventional wisdom is that large $N_c$ QCD has a Hagedorn spectrum: the number of hadrons with mass less than $m$ (denoted $N(m)$) in large $N_c$ QCD satisfies the condition that for sufficiently large $m$
\begin{equation}
 N(m) \ge e^{m/T}  \label{Hagspec}
\end{equation}
$T_H$ is defined as the maximum value of $T$ for which inequality~(\ref{Hagspec}) holds.  It {\it is} known that large $N_c$ QCD with adjoint fermions in 1+1 dimensions has a Hagedorn spectrum\cite{Kogan}.  While there is no real reason to doubt a Hagedorn spectrum also holds for large N$_c$ QCD in 3+1 dimensions, to date there has been remarkably little direct evidence for this.  There has been a study of the large $N_c$ glueball spectrum  based on a numerical treatment of a transverse lattice in a light cone formalism\cite{spec1}.  The results were consistent with a Hagedorn spectrum--but not definitive.   Indirect evidence for a Hagedorn spectrum at large $N_c$ can also be obtained  from lattice studies of thermodynamic properties of hadron matter in the metastable above the deconfining transition \cite{meta}.

The purpose of the present paper is to sketch a first principles argument that a Hagedorn spectrum must arise in QCD ({\it i.e.,} Yang-Mills fields plus quarks in the fundamental representation) at large $N_c$.  The approach relies only on generally accepted properties of QCD such as  asymptotic freedom  and confinement (in its basic sense that all physical states are color singlets) and on plausible assumptions about the applicability of perturbation theory to describe correlation functions at short distances.  It explicitly assumes neither that confinement is manifest through an unbroken center symmetry nor that the dynamics of the hadron is stringy.  The approach is based on the fact that the number of independent local operators with fixed quantum numbers and a single color trace grows exponentially with the mass dimension of the operator.  This approach is similar in spirit to the demonstration by Kogan and Zhitnitsky  that large $N_c$ QCD with adjoint fermions in 1+1 dimensions has a Hagedorn spectrum \cite{Kogan}. It also has elements which are reminiscent of refs.~\cite{smallV1,smallV2}; these works deal with a rather different problem, namely, the thermodynamics of systems confined to a small sphere rather than spectroscopy.

The approach relies on a confrontation between asymptotic freedom with confinement and depends on working in a regime in which perturbative corrections to the leading free-field values for certain correlation functions is small.  At first glance it may seem to be impossible to learn anything about the hadronic spectrum directly from the perturbative regime: the perturbative regime by construction only describes the smooth part of spectral functions.  However, this is misleading.  While one cannot learn anything directly about individual hadrons, one can learn certain global features about the spectrum, at least at large $N_c$.  Recall that in the formal large $N_c$ limit mesons are narrow.  High-lying mesons do not ``melt'' into the continuum due to large widths associated with phase space for decay.  Thus the spectral functions for correlators with meson quantum numbers are given entirely by mesons poles up to arbitrarily high mass.  The perturbative regime is for the deep space-like region of momentum space far from these poles; there is no ability to resolve information about individual hadrons in this regime. However, dispersion relations relate the space-like correlators to integrals over the spectral functions which are given by the meson poles.  Thus, while the perturbative regime cannot tell us about individual hadrons, it does contain some global information about the high-lying spectrum at large $N_c$.  As will be shown in this paper, the existence of a Hagedorn spectrum is global information about the spectrum which is extractable, provided one accepts certain technical assumptions about the applicability of perturbation theory.

Before turning to the Hagedorn spectrum, it is useful to recall a classic example where the nature of the asymptotically free regime for correlators is used to infer basic properties of the high-lying spectrum:  the demonstration  that large $N_c$ QCD has an infinite number of mesons with any given quantum numbers ({\it eg.,} the scalar-isoscalars) \cite{largeN2}.  Witten's original version of this argument was based on momentum space correlators.  Here it will be slightly recast in terms of correlators in (Euclidean) position space. This is useful since the argument for the existence of a Hagedorn spectrum is based on position space correlators.

Consider the correlator of the scalar current $J=\overline{q} q$ at two different space-times points.  By the standard K\"alle\'n-Lehmann spectral representation\cite{LehmannRep} it can always be written as
\begin{equation}
\begin{split}
&\Pi(r) \equiv \langle J(\vec{x}) J(0) \rangle = \int d s \rho(s) \Delta(r;s)  \\ & {\rm with} \; \; \Delta(r;s)=\int \frac{d^4 q}{(2 \pi)^4} \, \frac{e^{i \vec{q} \cdot \vec{x}} }{q^2 + s}=\frac{-1}{2 \pi^2 r} \frac{\partial K_0(\sqrt{s} r)}{\partial r}
\end{split}
\label{Pi1}\end{equation}
where $r=\sqrt{x^2+t^2}$ and $\rho(s)$ is the spectral function.
Large $N_c$ planarity implies that all physical states created by the current is composed of quarks and gluons in a single indivisible color singlet combination (plus $1/N_c$ corrections)\cite{largeN1,largeN2}.  If one imposes on this structure confinement (in the very basic sense that all physical states are color singlets), then one deduces that at large $N_c$ the currents can only makes single meson states\cite{largeN1,largeN2}.
The  spectral function at large $N_c$ is thus given by
\begin{equation}
\rho(s) = \sum_j |c_j|^2 \delta (s-m_j^2)
\label{rho}\end{equation}
where $m_j$ is the mass of  the $j^{\rm th}$ meson.

The scalar propagator, $\Delta$, has the property that
\begin{equation}
 \Delta (r,s)\rightarrow \frac{1}{2 \pi^2 r^2} \; \; {\rm as} \; \; r \rightarrow 0
\label{props}\end{equation} for any $s$; this asymptote is approached when $r \sqrt{s} \ll 1$.  Now suppose there were only a finite number of scalar-isoscalar mesons, then there must be a meson with a maximum mass.  In that case, Eqs.~(\ref{Pi1}) and (\ref{rho}) together with  Eq.~(\ref{props}) implies that
\begin{equation}
\Pi(r) \rightarrow \sum_{j=1}^{j_{max}} |c_j|^2/{2 \pi^2 r^2}
\label{counterfac}\end{equation}
for $r \ll m_{j_{max}}^{-1}$.   However, asymptotic freedom implies that when $r \ll \Lambda^{-1}$ (where $\Lambda$ is the QCD scale), $\Pi \sim r^{-6}$. This is inconsistent with the $r^{-2}$  scaling of Eq.~(\ref{counterfac}).  Thus we conclude that the assumption that there are only a finite number of mesons used to derive Eq.~(\ref{counterfac}) must be false: there is an infinite number of mesons at large $N_c$.

A couple of comments about this example are in order. first, the example shows that knowledge of correlation functions in the perturbative region is sufficient to extract some qualitative information about the high-lying spectrum of mesons for large $N_c$ QCD. The purpose of this paper is to show that the existence of a Hagedorn is similarly a qualitative property of the high-lying spectrum which is accessible from information in the perturbative regime.  second, it should be apparent that the procedure used in this example depended critically on taking $N_c \rightarrow \infty$ prior to the small $r$ limit.  It was only by taking $N_c \rightarrow \infty$ at the outset that one could justify the use of Eq.~(\ref{rho}) for small $r$.  In what follows, it  will always be assumed that the large $N_c$ limit is taken prior to any others.

The approach to establishing Eq.~(\ref{Hagspec}) is to generalize $J$ from a single operator to large {\it sets} of linearly independent  local operators all with the quantum number of interest.  Rather than having a single correlator as in the preceding example, one studies a matrix of correlators based on a set of local operators.  

The critical fact underlying this approach is the fact  that the number of distinct operators with fixed quantum numbers grows exponentially in the dimension of the operator.  This is easily shown.  Ultimately, this will translate into the condition that the number of  distinct hadrons at large $N_c$ must grow exponentially with the mass provided that the regime of validity  of perturbation theory is as expected.  Note that while there is a connection in this derivation between the exponential growth of the number of distinct operators with dimension and distinct hadrons with mass,  the connection is somewhat subtle.  The derivation does {\it not} assume that each operator couples to distinct hadronic states; and indeed they do not.  Instead the derivation exploits an inequality relating the lowest mass hadrons states with fixed quantum numbers to properties of the trace of the  logarithm of the correlator matrix.  

The strategy for deriving a Hagedorn spectrum has four basic components.  The first is based on the fact that as the spectral functions for the correlator matrix are saturated by meson poles at large $N_c$ (a fact also used to prove that there were an infinite numer of mesons at large $N_c$).  From this, plus general properties of the scalar propagator,  one can derive an inequality bounding the average of the lowest $k$ meson masses (where $k$ is the number of operators) from above by the derivative (with respect to $r$) of the average diagonal element of the logarithm of the matrix of correlators.

The second is a generalization from a single set of operators to an infinite sequence of sets of operators: ${\cal S}_1, \, {\cal S}_2, \, \ldots {\cal S}_n, \, \ldots$ . The key thing is that one can show that the inequality mentioned above implies a Hagedorn spectrum provided the dimension of the set of operators grows exponentially in  $n$ (the index specifying the term in the sequence) ({\it i.e.,} $\log(k) \sim n$), while the derivative (with respect to $r$) of the average diagonal element of the logarithm of the correlator matrix for the set of operator  asymptotes (at small $r$) to a constant times $-n/r$  with a correction whose relative size is bounded in a particular way.

The third ingredient is an explicit construction of such a sequence of sets of operators which one expects to meet the above criteria.  Here the focus will be on operators with the quantum numbers of scalar and pseudoscalar mesons.  The construction begins with local gluon bilinear operators coupled to a Lorentz scalar but in the color adjoint.  There are two such operators; one with  positive parity, the other with negative parity.  The $n^{\rm th}$ element of the sequence is the set of all single-color-trace operators composed of an anti-quark operator followed by $n$ of these bilinears followed by a quark operator.  This set clearly grows exponentially as there are $2^n$ elements.  Moreover, at asymptotically small $r$  and large $n$ asymptotic freedom requires that derivative of  the average diagonal element of the logarithm of correlator matrix asymptotes to $-8 n/r$, satisfying two of the conditions needed to demonstrate a Hagedorn spectrum.

The final step is to show that the  correction to $8 n/r$ has a relative size that is bounded appropriately.  At this point it is necessary to rely on perturbation theory.  It can be shown that perturbative corrections at any fixed order satisfy the requirement.  The critical point is to show that the correction {\it at  any given order in perturbation theory} has a contribution independent of $n$, plus $1/n$ corrections but no terms which grow as $n$.  Thus, to the extent that perturbation theory is valid---as is generally expected for correlators at short distance---a Hagedorn spectrum must emerge.

It is clear from this brief description, that argument does not constitute a rigorous theorem.  The reliance on perturbation theory is problematic.  In the first place it is approximate and thus difficult to use setting a precise bound.  second, it is an asymptotic series.  Due to the asymptotic nature of the series it is very difficult to see how to tighten this argument into a theorem.  The usual assumption in QCD is that  perturbation theory announces its own demise.  That is to say that if one studies correlation functions  beginning in the asymptotically free regime of short distances or high momenta, and then pushes to longer distances or lower momenta perturbative corrections (calculated at some order in perturbation theory) grow.  To the extent that these corrections remain a small fraction of the total, the usual expectation  is that perturbation theory should be valid.  The derivation here depends on this expectation holding for the matrix of correlation functions in question.  While it should be stressed that this is an {\it assumption}  and has not been proven rigorously, it should also be stressed that this assumption is quite standard.    To the extent that this expectation is correct, a ``physicist's proof'' of a Hagedorn spectrum might be said to exist.

The fact that argument is less than completely rigorous is hardly surprising---virtually nothing about QCD is known with full mathematical rigor including such basic features as asymptotic freedom which is also demonstrated perturbatively.  However, the  argument is of value for several reasons, despite its lack of rigor.  First, there is considerable experience that perturbation theory does accurately describe short-distance correlators; to the extent that this is true the Hagedorn spectrum has been established. Second, the argument can be easily generalized to various types of QCD-like theories with different numbers and types of fermions, different dimensions of space-time and different gauge groups.  The approach shows which of these have Hagedorn spectra and why.   Third, the argument does not explicitly assume that QCD approaches some type of a string theory for highly excited states.  To the extent that one takes a Hagedorn spectrum as a signature for stringy dynamics this argument gives insight into the emergence of stringy dynamics while only using generic properties of correlators.  Finally, the argument may provide some insight into the nature of confinement.  The argument only uses confinement in the sense that physical states are color singlets; but it does not explicit rely on confinement being manifest through an unbroken center-symmetry.

This paper is organized as follows: the next four sections will discuss each of the four basic components of the argument.  Following this will be a discussion on how the argument can be generalized to various QCD-like theories at large $N_c$ and some closing remarks.

\section{A useful Lemma}

The first step in establishing Eq.~(\ref{Hagspec}) is to generalize $J$ from a single current to large {\it sets} of linearly independent local gauge invariant operators constructed from quarks and gluons.  These operators will all have fixed quantum numbers of interest and  all involve only a single color trace.  This last condition ensures that at large $N_c$ the operator, when acting on the vacuum only creates single hadrons.  For technical simplicity here, these operators will be restricted to those with the quantum numbers of scalar-isoscalar mesons (where scalar refers to spin but not parity; pseudoscalars are included).  It is straightforward to generalize the argument to show exponential spectra for mesons of other quantum numbers and to glueballs, but the argument is particularly straightforward for scalar mesons.

In what follows, an arbitrary constant with dimensions of mass, $\Lambda$ is introduced and inserted in appropriate places so that correlation functions are dimensionless.  The precise value of $\Lambda$ is irrelevant as $\Lambda$ cancels out in all final results.

Begin with a set of $k$ linearly independent currents, single color trace, scalar-isoscalar currents ${\cal S}= \{J_1, J_2, J_3 \cdots J_k \}$. Associated with $\cal{S}$ is a dimensionless  $k \times k$  mixed correlator matrix $\overleftrightarrow{\Pi}^{\cal S}(r)$.  For any operators  $J_a,J_b \in {\cal S}$ the matrix elements are given by
\begin{equation}
\Pi_{a b}(r) \equiv \frac{ \langle J_a^{\dagger}(\vec{x}) J_b(0) \rangle}{\Lambda^{d_a+d_b}} = \int d s \rho_{ a b}(s) \Delta(r;s)  \;\;  {\rm with} \; \rho_{a b}(s)=\sum_{j=1}^\infty c_{a,j}^* c_{b,j} \delta (s- m_j^2) ;
\label{Pi}
\end{equation}
 $J_a^{\dagger}$ is the Hermitian adjoint of $J_a$ , $d_c$ is the mass dimensions of  operator $J_c$  and  $c_{a,j}$ is the (dimensionless) amplitude for the $a^{\rm th}$ current to create the $j^{\rm th}$ scalar-isoscalar meson:  $c_{a,j} \equiv \Lambda^{-d_a} \langle j|J_a|{\rm vac} \rangle$ (where $|j \rangle$ is the hadronic state associated  with the $j^{\rm th}$ scalar-isoscalar meson).  The form of the spectral function follows from the usual large $N_c$ requirements.

Given a set of such operators, ${\cal S}$,  an important lemma can be established for meson masses in the large $N_c$ limit.
The lemma is that for any $r>0$,
\begin{equation}
\overline{M}_{\cal S}(r)\equiv -\, \frac{d}{d r} \, \frac{ {\rm tr} \left (  \log \left (\overleftrightarrow{\Pi}^{\cal S}(r) \right ) \right )}{||{\cal S}|| }  \ge \, \sum_{j=1}^{||{\cal S}||}\frac{ m_j}{{||{\cal S}||}} \, \, ;
\label{lemma}
\end{equation}
$m_1$ is the lightest scalar-isoscalar meson, $m_2$ the second lightest, {\it etc.,}~and $||{\cal S}||= {\rm dim}\left ( \overleftrightarrow{\Pi}^{\cal S}(r) \right ) =k$.  Note that this lemma bounds the average of the lowest $k$ mesons by minus the derivative of  the average diagonal matrix element of the logarithm of the correlator matrix.

This lemma can be established from general properties of linear algebra,  Eq.~(\ref{Pi}) and along with a few basic properties of the scalar propagator function as discussed in Appendix \ref{A1}.

\section{A sequence of sets of operators}

Consider a {\it sequence} of sets of scalar-isoscalar linearly independent gauge invariant local operators  with quantum numbers of a scalar-isoscalar meson and composed of single color trace operators, ${\cal S}_1, {\cal S}_2, {\cal S}_3,\ldots, {\cal S}_n, \ldots \;$ .  Suppose that such a sequence exists with the property that one can always find positive constants $r_0$, $A$, $p$, and $n_0$ such that for all $r<r_0$  and all $n>n_0$ the following two conditions are satisfied:
\begin{equation}
 a)  \; \;||S_n|| \ge A^n   \; \; ; \; \; \;b)  \; \; n \frac{p }{r} \ge  \overline{M}_{{\cal S}_n}(r) \; .
\label{seqprop}\end{equation}
Given such a sequence one can establish the Hagedorn spectrum of Eq.~(\ref{Hagspec}).

To see this consider $N_0(m)$, the number of scalar-isoscalar mesons with mass less than $m$.  By construction $N(m)$, the total number of hadrons with mass less then $m$, is greater then or equal to $ N_0(m)$.  Define
\begin{equation}
\langle m \rangle_0 \equiv \frac{\sum_j m_j \theta (m_0-m_j)}{N_0(m_0)}
\end{equation}
where $m_j$ is the mass of the $j_{\rm th}$ scalar meson; {\it i.e.},
$\langle m \rangle_0$ is the average mass for all scalar mesons with mass less than $m_0$.   Start by assuming that $N_0(m)$ grows with $m$ at least linearly.  In that case $\langle m \rangle_0 \ge m_0/2$.  Choose a value of $m_0$ such that $N_0(m_0)=||{\cal S}_n||$.  Using the lemma in Eq.~(\ref{lemma}) and condition b) in Eq.~(\ref{seqprop}) allows one to deduce that
\begin{equation}
 n \frac{p }{r_0} \ge \overline{M}_{{\cal S}_n} \ge \frac{1}{||{\cal S}_n||} \sum_{j=1}^{||{\cal S}_n||} m_j \ge \langle m \rangle_0 \ge \frac{m_0}{2} \; .
\end{equation}
Thus $2 n p/r_0\ge m_0$.  Acting on both sides of this with the monotonic function $N_0$ yields
\begin{equation}
N_0(2 n p/r_0)\ge N_0(m_0)=||{\cal S}_n|| \ge A^n
\end{equation}
where the last inequality follows from condition a) of  Eq.~(\ref{seqprop}).  Finally, denoting $ 2 n p/r_0$ as $m$ yields
\begin{equation}
N(m) \ge N_0(m) \ge (A^{{r_0}/{2 p}})^m = e^{m ({r_0 \log(A)}/{2p} )}
\label{Hagf}\end{equation}
This is of the form of a Hagedorn spectrum with
\begin{equation}
T_H \le \frac{2 p}{r_0 \log (A)}
\label{TH} \end{equation}

This derivation depended on the assumption that $N_0(m)$ scaled at least  linearly.  This assumption is, of course, justified {\it a posteriori} by the form of Eq.~(\ref{Hagf}).  While the preceding argument is sufficient to obtain a Hagedorn spectrum, it is worth noting that the bound on $T_H$ in Eq.~(\ref{TH}) can be sharpened.  Note that the factor of $2$ came from the assumption that $N_0(m)$ grew at least linearly.  If instead the assumption had been that $N_0$ grew at least as fast as a given power law: $N(m) \ge {\rm const} \times  N^\alpha$ (which is also justified {\it a posteriori}) then the condition on $T_H$ becomes $T_H \le \frac{(1 +\alpha) p}{\alpha \, r_0 \log (A)}$ for all $\alpha$.  Taking $\alpha$ to be arbitrarily large yields
\begin{equation}
T_H \le \frac{ p}{r_0 \log (A)} \; .
\label{TH2} \end{equation}

\section{Constructing  operators}

The next step is to construct explicitly a sequence of sets of operators which satisfy the conditions in Eq.~(\ref{seqprop}) and thereby establish a Hagedorn spectrum.

Two useful ingredients in this construction are the scalar and pseudoscalar operators composed of bilinears in the gluon fields.
\begin{equation}
{\cal O}_+ \equiv \frac{1}{N_c^2} F_{\mu \nu} F^{\mu \nu} \; \; \;  {\cal O}_- \equiv \frac{1}{N_c^2} F_{\mu \nu} \tilde{F}^{\mu \nu} \; .
\end{equation}
These operators are {\it not} traced over color; and at large $N_c$ they become pure color adjoint operators up to $1/N_c$ corrections.   From these one can construct operators of the following form
\begin{equation}
 J_{l_1,l_2,\cdots, l_n} \equiv \overline{q}\, {\cal O}_{l_1} {\cal O}_{l_2} \cdots  {\cal O}_{l_n} q \; \; \;{\rm where} \; \;{l_1, \cdots ,  l_n}= \pm
\label{ops}
\end{equation}
{\it eg.}, $J_{+-}=\overline{q}\, {\cal O}_+ {\cal O}_-q= \overline{q} F_{\mu \nu} F^{\mu \nu} F_{\alpha  \beta} \tilde{F}^{\alpha \beta} q/N_c^4$.  Define ${\cal S}_n$ to be the set of all operators of this form of (engineering) dimension $4n+3$, {\it i.e.}, containing $n$ ${\cal O}_l$ operators. Thus,
\begin{equation}
\begin{split}
&{\cal S}_1=\{ J_+, J_- \} = \left \{\frac{\overline{q} F_{\mu \nu} F^{\mu \nu}  q}{N_c^2} , \frac{\overline{q} F_{\mu \nu} \tilde{F}^{\mu \nu} q }{N_c^2}\right \} \\
&{\cal S}_2=\{ J_{++}, J_{+-}, J_{-+},J_{--} \}= \left \{ \frac{\overline{q} F_{\mu \nu} F^{\mu \nu}  F_{\alpha \beta} F^{\alpha \beta}q}{N_c^4}, \cdots \right \} \\
& {\cal S}_3=\{ J_{+++}, J_{++-}, J_{+-+}, J_{+--}, J_{-++},J_{-+-}, \cdots \} \\
& \cdots
\end{split}
\end{equation}
 By construction $||{\cal S}_n||=2^n$ so that the sequence  ${\cal S}_1, {\cal S}_2, {\cal S}_3,\ldots, {\cal S}_n, \ldots$ obviously satisfies condition a) of Eq.~(\ref{seqprop}); demonstrating that condition b) holds is equivalent to establishing a Hagedorn spectrum.

\section{Establishing the conditions needed for a Hagedorn spectrum}

\begin{figure}
\centering
\includegraphics[width=3.25in]{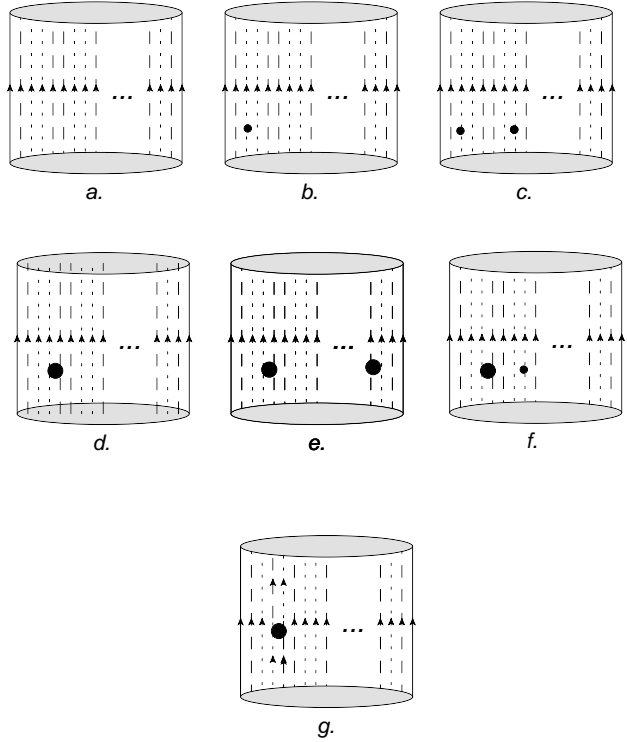}
\caption{Typical diagrams which contribute to diagonal correlators associated with the currents in ${\cal S}^{n}$.  The solid lines correspond to quark propagators.  The dotted lines (dashed lines) represent the free propagation of two gluons coupled to a scalar (pseudo-scalar) as created by ${\cal O}_+$ (${\cal O}_-$).  The black circles correspond to interactions due to one or more gluon exchanges.  The smaller black circles represent gluon exchanges within one pair of gluons while the larger circles represent gluon exchanges between two pairs.    The ellipsis indicates pairs of gluons which are not explicitly represented. The large grey ovals correspond to operators in ${\cal S}^{n}$. In diagrams a.~through f.~the initial operator and final operators are the same; these diagrams contribute to diagonal matrix element in the matrix of correlators.  In diagram g.~the initial and final operators differ and corresponds to a contribution to a mixed correlator, {\it i.e.} an off-diagonal matrix element. }

 \label{hagfig}
\end{figure}

To establish condition b) of Eq.~(\ref{seqprop}) for this sequence of sets of propagators---and thereby establish the existence of a Hagedorn spectrum---one begins with an analysis of the short-distance behavior of $\overleftrightarrow{\Pi}(r)$.  As $r \rightarrow 0$, asymptotic freedom  implies that correlators take their free-field values (as represented for example by diagram a.~of Fig.~\ref{hagfig}). Since the two-gluon state created by ${\cal O}_+$ associated with some particular color indices are free fields they cannot be annihilated by ${\cal O}_-$.  Moreover, the chance that the same color index appears more than once among the  ${\cal O}_\pm$ operators vanishes at large $N_c$. Thus at large $N_c$ and $r \rightarrow 0$ the correlator will vanish unless all of the  ${\cal O}_\pm$ operators in each current occurs in the same order: $\overleftrightarrow{\Pi}$ is diagonal.

Moreover when $r$ is much smaller than $\Lambda_{\rm QCD}^{-1}$, simple dimensional analysis dictates the form of the diagonal matrix elements of $\overleftrightarrow{\Pi}$; they must all be of the form ${\rm const}\,  r^{-(8 n +6)}$ since $\overleftrightarrow{\Pi}$ is of mass dimension $8 n +6$ and $r$ is the only dimensional parameter left at those scales.  From this it is easy to see that $\overline{M}_{{\cal S}_n}(r) \rightarrow 8 n + 6$  as $ r \rightarrow 0 $.
Given this  asymptotic form, it is useful to parameterize $\overline{M}_{{\cal S}_n}(r)$ in the follow way
\begin{equation}
\overline{M}_{{\cal S}_n}(r) = n \frac{8 + \frac{6}{n}}{r }  + R(r,n)
\label{R1}\end{equation}
where $R(r,n)$ is the size of the correction due to interactions.  The condition for a Hagedorn spectrum amounts to the condition that there exists a number $\rho$ such that
\begin{equation}
R(n,r) < n \frac{\rho}{r} \; \;  {\rm whenever}  \;  n >n_0  \; {\rm and} \; r <r_0 \; ;
\label{rho}\end{equation}
in that case condition b) of Eq.~(\ref{seqprop}) is satisfied with $p=8 + 6/n_0+\rho$.

To study the behavior of $R(r,n)$, it is useful to parameterize $\overleftrightarrow{\Pi}$ as the product of its free field value times a correction due to interactions:
\begin{equation}
\begin{split}
&\overleftrightarrow{\Pi}_{{\cal S}_n}(r) \equiv \overleftrightarrow{\Pi}_{\rm free} (r) \left ( \overleftrightarrow{1} + \overleftrightarrow{C}_{{\cal S}_n}(r) \right ) \\
& \log \left (\overleftrightarrow{\Pi}(r) \right ) =
\log \left ( \overleftrightarrow{\Pi}_{\rm free}(r) \right )
- \sum_{j \ge 1} \frac{(-1)^j}{j} \overleftrightarrow{C}^j(r)
\label{C}
\end{split}
\end{equation}
From the definition of $\overline{M}_{{\cal S}_n}(r)$  and $R$ one sees that
\begin{equation}
  R(r,n) = \partial_r \left( \sum_{j \ge 1} \frac{(-1)^j}{j} {\rm tr} \left( \overleftrightarrow{C}^j(r) \right ) \right)
\label{R2}
\end{equation}

Ideally one should obtain a rigorous bound on $R(r,n)$.  It is not immediately apparent how to do this.  However, given that the objects under study are correlation functions, it is natural to make the standard assumptions used in the study of QCD correlation functions---namely, that at short distances correlators are accurately described by renormalization-group-improved perturbation theory. Under this assumption, it will be shown that condition b) holds {\it when working up to any given order in perturbation theory}; it will then be argued that non-perturbative effects are not expected to alter this conclusion.  Before proceeding with a perturbative analysis it is worth noting that this result is nontrivial.  It is not sufficient to argue that at short distances perturbative corrections generically become small since $\alpha_s$ becomes small.  The key issue is that generically, the coefficients in a perturbative expansion ({\it i.e.}, an expansion in $\alpha_s$) for $R$ can be expected to depend on $n$.  If any of these coefficients grow with $n$ faster than linearly then condition b) of Eq.~(\ref{seqprop}) is not satisfied for the perturbative estimate of $\overline{M}_{{\cal S}_n}$.  Moreover, the typical behavior of combinatoric factors in perturbative theory might lead one to expect that coefficients would diverge with $n$.  However by taking the large $N_c$ limit at the outset, these combinatoric factors are greatly suppressed, yielding a linear dependence on $n$.

Suppose the correlators in the set of  ${\cal S}_n$ (specified above) are computed perturbatively.  The expansion at large $N_c$ is in the `t Hooft coupling $N_c \alpha_s$.  Suppose one works to some fixed order,  $(N_c \alpha_s)^l$, where $\alpha_s$ is taken to be evaluated at the scale $\mu^2=r^{-2}$.  It can then be shown  for $n \ge l+1$ that
\begin{equation}
\begin{split}
R(r,n_0) = \partial_r \left (
 n \sum_{i=1}^l \, (N_c \alpha_s)^i \,  g_i
+ \sum_{i=1}^l \, (N_c \alpha_s)^i \,  h_i \right )
\end{split}\label{Trform}\end{equation}
where $g_i$ and $h_i$ are numerical coefficients.  Note that the only $r$ dependence on the right-hand side is through the scale in $\alpha_s$.   The two essential features  of Eq.~(\ref{Trform}) are that the coefficients $g_i$  and $h_i$ are {\it universal}---their values do not depend on $n$---and that  at large $n$, the dominant contributions  scale with $n$ but not faster.

The derivation of Eq.~(\ref{Trform}) is somewhat involved.  However, the origins of its key features are relatively easy to understand.   One critical ingredient  is the use of correlators in position space. This ensures that contributions of non-interacting parts of a correlator simply factorize:  the contribution to a given diagram of a particular type of cluster involving the interaction of some number of  ${\cal O}_\pm$ currents is independent of $n$. This gives rise to the universality of the coefficients $g_i, h_i$.  A second key ingredient is the logarithmic structure of $\overline{M}_{{\cal S}_n}$.

The final critical ingredient in obtaining Eq.~(\ref{Trform}) is the large $N_c$ limit.  This suppresses non-planar diagrams and thus  implies that clusters of interacting gluons only involve neighboring ${\cal O}_\pm$ currents.  This limits the combinatoric growth of the amplitudes with $n$---ultimately restricting it no faster than $n$.   To illustrate this, consider one gluon exchange contributions to some diagonal matrix element of $\overleftrightarrow{C}$.  The  Feynman diagram contains 2 quark lines and 2-n gluons which enter and leave the operators. Thus, at the  one-gluon exchange level there are $(2 n + 2 )^2$ distinct one-gluon exchange contributions between these lines.  This grows faster than $n$.  However, most of these gluon exchanges are non-planar and thus suppressed at large $N_c$.  There are only $2n+1$ planar contributions and this growth is linear in $n$---as is required.

The logarithmic structure of $\overline{M}_{{\cal S}_n}$ plays an  essential role in restricting the scaling of $R(r,n)$  with $n$ to no faster than linearly.  To see why, consider diagrams b.~ and c.~of Fig.~\ref{hagfig}.

Diagram b.~of Fig.~\ref{hagfig} represents the  contribution  of all gluon exchanges up to fixed order $j$ in perturbation theory contained within one the positive parity pair of gluons.  In effect it is a self-energy contribution for the pair as a whole---for one pair in one diagonal matrix element in $\overleftrightarrow{C}(r)$.  Denote this contribution to the diagonal matrix element as $\Sigma_+$.  Note that $\Sigma_+(r)$ can be calculated perturbatively and expressed as a series in $N_c \alpha_s$.    Next note that because the correlator is point-to-point, contributions factorize.  Thus one has the same self-energy contribution to this matrix element from any positive parity pair; the total contribution to the matrix element from all diagrams with a single self-energy of a positive parity pair is $n_+ \Sigma_+$, where $n_+$ is the number of positive parity pairs in the operator $J_a$.  The contribution of all such diagrams to ${\rm tr} \left( \overleftrightarrow{C}(r) \right )$ is thus $\langle n_+ \rangle \Sigma_+$ (where the averaging is over states in the class) yielding a total contribution of $n \Sigma_+/2$; the contribution to $R(r,n)$ from the $j=1$ term in Eq.~(\ref{R2}) from these diagrams is $n \partial_r \Sigma_+/2$.  As expected this contribution to $R(r,n)$  grows linearly with $n$.

However, there are also contributions which scale as $n^2$.  First of all there is the $j=2$ term in Eq.~(\ref{R2}); this depends on ${\rm tr} \left( \overleftrightarrow{C}^2(r) \right )$.  The contributions to this term from diagram b.~acting twice is $-((n^2+n)/8) \partial_r \Sigma_+^2 $.  There is also a contribution from diagrams containing two distinct self-energies such as seen in diagram c~.  The $j=1$ contribution for these is $((n^2-n)/8) \partial_r \Sigma_+^2 $.  Note that this $n^2$ contribution exactly cancels the previous one yielding a total contribution to $R(r,n)$  proportional to $n$.  The reason for this cancelation is clear: the combinatoric factors are identical.

This behavior is generic.  Cancelations between separate interactions within diagrams and higher powers in $\overleftrightarrow{C}^2(r)$ always occur yielding a linear growth in $n$ and not faster.  It occurs for interactions involving clusters of adjacent currents.  An example is in diagram d.~of Fig.~\ref{hagfig}, which represents gluon exchanges which couple one negative parity pair of gluons with a positive parity pair.  Denoting the amplitude for this as $\Sigma_{-+}$ is easy to show that the contribution from all diagrams containing one such pair to the $j=1$ contribution to $R(r,n)$ is $(n-1))\Sigma_{-+}/4$.  The contribution of order $n^2$ from $j=2$ from diagrams of type b.~and d.~cancel against the order $n^2$ contributions from the $j=1$ contributions in diagrams of type c.,~e., and f.  Again, the reason for this is that the combinatoric factors are the same.  This general behavior has been explicitly checked for a large number of cases and appears to hold for diagrams containing any number of clusters and for clusters of any size; in all cases cancelations occur in which the leading behavior after cancelations scales as $n$ but not higher.  Moreover, it was also explicitly checked to hold for off-diagonal matrix elements such as seen in  diagram g.~of Fig.~\ref{hagfig}; these off-diagonal contributions only contribute for $j=2$ and higher.

Equation (\ref{Trform}) follows directly from these considerations.  Note the underlying amplitudes associated with clusters such as $\Sigma_+$ and $\Sigma_{-+}$ are computable perturbatively. Moreover,  the number of distinct cluster types which can contribute is fixed by the order of perturbation theory $l$: a cluster of size $n$ requires at least $n-1$ gluon exchanges.  Thus when working at any fixed order $l$, one needs only amplitudes for clusters of size $l+1$ or less. It should also be clear why the coefficients in Eq.~(\ref{Trform}) for all $n>l$ are universal.

Next evaluate the derivative in Eq.~(\ref{Trform}).   The right-hand side only depends on $r$ through the running of the coupling, which is given by a large $N_c$ RG equation with a $\beta$ function calculable in perturbation theory:
\begin{equation}
\frac{d(N_c \alpha_s)}{d \, r} = r^{-1} \sum_{i = 1}^l \overline{b}_i {(N_c \alpha_s)}^i
\end{equation}
where the $\overline b_i$ are fixed coefficients.  To order $(N_c \alpha_s)^{l+1}$, $R(n,r)$ is accurately given by
\begin{equation} \begin{split}
&R(n,r) =  \frac{\sum_{i=2}^{l+1} \left( c_i  + \frac{d_i}{n} \right ) (N_c \alpha_s)^i  )}{r}\\
&c_i= \sum_{m=1}^{l} \sum_{m'=1}^{l} g_m \overline{b}_{m'} \delta_{m+m', i} \\
&d_i= \sum_{m=1}^{l} \sum_{m'=1}^{l} h_l \overline{b}_{m} \delta_{m+m', i}
\end{split}
\label{f}\end{equation}
where $g_i$ and $h_i$ are the coefficients in Eq.~(\ref{Trform}).
Note that $N_c \alpha_s(\mu^2)$ is a monotonically decreasing function and asymptotes to zero.  Thus, setting $\mu^2=1/r^2$  one can always find  a value $r_0$ such that $N_c \alpha_s$ evaluated at $\mu =1/r > 1/r_0$ can be made as small as one likes.   For sufficiently small $N_c \alpha_s$ the series is dominated by its first term ($i=2$) and then clearly satisfies Eq.~(\ref{rho}).  Thus  the last condition needed to establish a Hagedorn spectrum is satisfied provided standard assumptions about the regime of validity of perturbation theory apply.

\section{Discussion}

The argument of the previous four sections demonstrates that a Hagedorn spectrum must exist for large $N_c$ QCD provided that renormalization-group improved perturbation theory is applicable for correlation functions at sufficiently short distances.

One obvious drawback of this approach is its reliance on perturbation theory for the correlation functions.  Of course, there is a general expectation that perturbation theory will be accurate at short distances.   Moreover,  vacuum condensates yielding  power law corrections to perturbation theory  (as one has for example in the QCD sum rule approach to phenomenology\cite{SVZ}) can be incorporated in the present approach in a straightforward way.  The contributions of the condensate factorize in a manner analogous to the perturbative corrections and do not affect the conclusions.  Nonetheless, the reliance of perturbation theory raises  the issue that the perturbative expansion is asymptotic. This appears to preclude a straightforward method for strengthening into a rigorous theorem the heuristic reasoning used here.  However, despite its lack of rigor the approach does give a strong argument as to why a Hagedorn spectrum is expected without any reliance on the assumption of stringy dynamics.

The present approach can be generalized easily.  With modest changes, it can be used to show that  mesons and glueballs of any quantum number also have exponentially growing spectra.  The modification of the argument for mesons with quantum numbers other than scalar (and pseudoscalar) involves little more than altering the form of the dispersion relation to account for spin.  As in the case of spinless mesons, one can insert arbitrary numbers of ${\cal O}_{\pm}$ operators between the quark creation and annihilation operators without affecting the quantum numbers of the operator.  Thus the number of operators grows as $2^n$ and the argument goes through essentially unchanged.

There is a small subtlety for the case of glueballs.  Consider the case of scalar and pseudoscalar glueballs.  The natural class of  operators to pick are of the form
\begin{equation}
 J_{l_1,l_2,\cdots, l_n} \equiv {\rm Tr}\left( {\cal O}_{l_1} {\cal O}_{l_2} \cdots  {\cal O}_{l_n}  \right ) \; \; \;{\rm where} \; \;{l_1, \cdots ,  l_n}= \pm
\label{ops}
\end{equation}
and the trace is in color space.  Because of the cyclic property of the trace these operators are not all distinct. For example  $J_{++-}$ is identical to $J_{+-+}$.  This raises the obvious question of how many {\it distinct} operators exist for any fixed $n$.  As it happens, the combinatorics are nontrivial and there is no simple analytic formula for this for general $n$.  However, it is easy to show that when $n$ is a prime number the number of distinct operators is $(2^n -2n +1)/n$; when $n$ is not prime it is always larger than $(2^n-2n+1)/n$.  Thus, the number of operators still grows faster than any exponential with base less than 2 and the basic argument goes through. 
 
The argument also applies to other QCD-like theories at large $N_c$.  These include pure gauge theory,  theories with quarks in non-fundamental representations: the adjoint representation, the two-index symmetric representation or the two index antisymmetric representation \cite{Orienti}.  The latter may be of  interest for phenomenological reasons---at $N_c=3$ it coincides with QCD.  The approach also applies to large $N_c$ gauge theory (with and without matter fields) for other gauge groups ({\it eg.,} $SO(N_C)$).

All of the systems with Hagedorn spectra considered above have an unbroken effective center symmetry at large $N_c$. This interesting since an unbroken center symmetry is often taken as a signature of confinement.   In cases such as pure gauge theory or theories with adjoint fermions, the center symmetry is manifest.  For the other cases it is an emergent symmetry---while not being a symmetry of the theory at finite $N_c$, the effects of explicit center symmetry breaking becomes small as $N_c$ increases and goes to zero at infinite $N_c$.

For QCD with fundamental quarks, center symmetry trivially emerges at large $N_c$ since quark loops are suppressed while the pure Yang-Mills sector has an explicit center symmetry. The emergence of an effective center symmetry at large $N_c$ depends on nontrivial large $N_c$ dynamics in other cases. For example, in theories with quarks in the  anti-symmetric two-label representation\cite{center},  the emergent center symmetry may be thought of as due to an orientifold equivalence at large $N_c$ between this and QCD with adjoint matter.  Other cases can be seen to have an emergent symmetry due to orbifold considerations\cite{orbi}.  It is interesting that although all these cases have an emergent center symmetry, the derivation of the Hagedorn spectrum did not rely explicitly on confinement in the sense of an unbroken center symmetry; it only used confinement in the sense of an absence of non-singlet physical states.  The question of whether an unbroken center symmetry is {\it required} for the emergence of a Hagedorn spectrum will be explored in future work.

The author thanks  D.T. Son, L.Ya. Glozman, A. Cherman and M. Shifman for useful discussions.  He is indebted to R. Pisarski  for pointing out the connections to refs. \cite{smallV1,smallV2}, and to Michael Cohen for an explanation of the combinatorial aspects of the glueball operators.  The support of the U.S. Department of Energy is gratefully acknowledged.

\appendix
\section{The lemma \label{A1}}

 Before proving the lemma of Eq.~(\ref{lemma}), it is worth remarking that the logic underlying it is quite standard in the lattice QCD community.  It was realized long ago that the masses of  the lowest lying states coupling to a set of currents with fixed quantum numbers are extractable as the large $r$  limit of the derivative of the eigenvalues of $ \log \left (\overleftrightarrow{\Pi}^{\cal S}(r) \right )$.   Indeed attempts to extract excited hadron masses from numerical lattice studies routinely use this approach\cite{lattice}. The lemma follows simply by averaging over all the masses extractable this way  in principle while working at large $N_c$ (where the mass spectrum is discrete) and exploiting the following standard  properties of the propagator  which hold for positive values of $r$ and $s$.  
\begin{equation}
i) \Delta(r,s) >0  \: \; \; \; \; 
ii)-\frac{\partial^2 \log \left ( \Delta (r,s) \right ) }{\partial r^2} < 0 \; \; \; \; \;
 iii)  \; \;  -\frac{\partial \log \left ( \Delta (r,s) \right ) }{\partial r} \rightarrow \sqrt{s} \; \; {\rm as} \; \; r \rightarrow \infty  \; .
\label{prop}\end{equation} 
 
A formal proof of the lemma starts with an easily derived identity: if $\overleftrightarrow{H}(a)$ is a sufficiently smooth matrix valued function of invertible matrices then
\begin{equation} 
-\frac{d^2 \, {\rm tr} \left (  \log \left (\overleftrightarrow{H}(a) \right ) \right )} {d a^2} = {\rm tr}  \left ( \overleftrightarrow{H}(a)^{-1} \overleftrightarrow{H}'(a) \overleftrightarrow{H}(a) ^{-1} \overleftrightarrow{H}'(a)\right )  - {\rm tr} \left ( \overleftrightarrow{H}(a)^{-1} \overleftrightarrow{H}''(a) \right) 
\label{Ident}\end{equation}
where the prime indicates differentiation.   Suppose  that $\overleftrightarrow{H}(a)$ is  Hermitian and positive definite for all finite positive $a$ and that   $\overleftrightarrow{H}''(a)$ is negative definite.  Then both traces the right-hand side of Eq.~(\ref{Ident}) can easily be seen to be necessarily positive.    This implies that for  $a>0$ ,   $-\frac{d}{ d a}  \, {\rm tr} \left (  \log \left (\overleftrightarrow{H}(a) \right ) \right )   > \lim_{a \rightarrow \infty}  \left ( -\frac{d}{ d a}  \, {\rm tr} \left (  \log \left (\overleftrightarrow{H}(a) \right ) \right )  \right ) $.   Note that $\overleftrightarrow{\Pi}^{\cal S}(r) $  is Hermitian.  Thus,  providing that $\overleftrightarrow{\Pi}^{\cal S}(r) $ is positive definite and ${\partial_r^2}\left ( \overleftrightarrow{\Pi}^{\cal S}(r) \right) $  is negative definite,
\begin{equation} 
-\frac{d}{ d r}  \,\frac{{\rm tr} \left (  \log \left (\overleftrightarrow{\Pi}^{\cal S}(r) \right ) \right )} {||{\cal S}||}  > \lim_{r \rightarrow \infty}  \left ( -\frac{d}{d r} \, \frac{ {\rm tr} \left (  \log \left (\overleftrightarrow{\Pi}^{\cal S}(r) \right ) \right )}  {||{\cal S}||}  \right ) \; .  \label{Piineq}\end{equation}

The demonstration that $\overleftrightarrow{\Pi}^{\cal S}(r) $ is positive definite and ${\partial_r^2 }\left ( \overleftrightarrow{\Pi}^{\cal S}(r)  \right) $  is negative definite  follows  from the definition of  $\overleftrightarrow{\Pi}^{\cal S}(r) $ in Eq.~(\ref{Pi}) which allows one to rewrite  $\overleftrightarrow{\Pi}^{\cal S}(r) $ as
\begin{equation}
 \overleftrightarrow{\Pi}^{\cal S}(r) = \sum_{j=1}^\infty \overleftrightarrow{\pi}^{(j) \cal S}\Delta(r,m_j^2)   \; \; \; {\rm with} \; \; \;  {\pi}_{a b}^{(j) \cal S}(r)=c_{a,j}^* c_{b,j}  \; ;
\label{Piform} \end{equation}
  the sum is over mesons.  By construction $\overleftrightarrow{\pi}^{(j) \cal S}$ is positive definite: it is a positive constant times a projection operator on to a single vector.  Combining this with  property i)  of  Eq.~(\ref{prop}) implies that  $\overleftrightarrow{\Pi}^{\cal S}(r) $ is positive definite.  Similarly, combining it with property ii) implies that ${\partial_r^2 }\left ( \overleftrightarrow{\Pi}^{\cal S}(r)  \right) $  is negative definite.

The final step is to show that 
\begin{equation}
\lim_{r \rightarrow \infty}  \left ( -\frac{d}{d r} \, \frac{ {\rm tr} \left (  \log \left (\overleftrightarrow{\Pi}^{\cal S}(r) \right ) \right )}  {||{\cal S}||}  \right )  \ge   \, \sum_{j=1}^{||{\cal S}||}\frac{ m_j}{{||{\cal S}||}} 
\label{finalgoal}\end{equation} which together with Eq.~(\ref{Piineq}) establishes the lemma.  To do so consider the sum in Eq.~(\ref{Piform}).   One can construct a new operator by truncating the sum over mesons:
\begin{equation}
 \overleftrightarrow{\Pi}_{\rm trunc}^{\cal S}(r) = \sum_{j \in {\cal H}} \overleftrightarrow{\pi}^{(j) \cal S}\Delta(r,m_j^2)   \; ,
\label{Piform2} \end{equation}
where the mesons included in this sum are in a set ${\cal H}$ with the following properties: i) $||{\cal H}||=||{\cal S}|| $ ({\it i.e.}, there is one state per operator in ${\cal S}$ ); ii) elements of all of the matrices $ \overleftrightarrow{\pi}^{(j) \cal S}$ for mesons in the set are linearly independent and nonzero; and iii)  the mesons in the set are the lightest ones possible consistent with previous conditions.  Generically one expects that in the absence of symmetries this set will simply be the lightest $||{\cal S}||$ mesons.  Condition iii) of Eq.~(\ref{prop}) implies that as $r \rightarrow \infty$ , $ \overleftrightarrow{\Pi}^{\cal S}(r)  \rightarrow \overleftrightarrow{\Pi}_{\rm trunc}^{\cal S}(r)$ since the effects of  the states not in ${\cal H}$ make exponentially suppressed contributions which vanish as $r \rightarrow \infty$.    The exponential suppression of higher mass states at large $r$ also greatly simplifies the computation of eigenvectors and eigenvalues of $ \overleftrightarrow{\Pi}^{\cal S}(r) $ at  large $r$.  Recall that up to a multiplicative constant the matrices  $\overleftrightarrow{\pi}^{(j) \cal S}$  are  projections on to a single vector $\vec{v}_j$.  The eigenvectors  of $ \overleftrightarrow{\Pi}_{\rm trunc}^{\cal S}(r)$ at large $r$ can be constructed from these.  Clearly $\vec{v}_1$ becomes an eigenvector at large $r$ since the contributions of all mesons in the set except the lightest makie a contribution which is  exponentially small compared to the leading one.  The associated eigenvector is $\Delta(r,m_1^2)  \sum_{k=1}^{||{\cal S}||} |c_{1,k}|^2$.  One obtains a second eigenvector  at large $r$ by starting with $\vec{v}_2$ and projecting out of $\vec{v}_1$.  Higher mass states exponentiate away while the contribution from the lightest state is removed by projection; the eigenvalue is $\Delta(r,m_2^2)  \sum_{k=1}^{||{\cal S}||} |c_{2,k}|^2$.  One can construct all eigenvectors in a similar manner; the approach is the essentially Gram-Schimdt procedure: the third eigenvector is $\vec{v}_3$ with $\vec{v}_2$ and $\vec{v}_1$  projected out and so forth.  The $j^{\rm th}$  eigenvalue is thus $\Delta(r,m_j^2)  \sum_{k=1}^{||{\cal S}||} |c_{j,k}|^2$   for all $j \in {\cal H}$.  Together with condition iii) of Eq.(\ref{prop}) this implies that 
\begin{equation}
\lim_{r \rightarrow \infty}  \left ( -\frac{d}{d r} \, \frac{ {\rm tr} \left (  \log \left (\overleftrightarrow{\Pi}_{\rm trunc}^{\cal S}(r) \right ) \right )}  {||{\cal S}||}  \right )  \ge   \, \sum_{j \in {\cal H}}\frac{ m_j}{{||{\cal S}||}} \; .
\label{last}\end{equation}
However,  $  \sum_{j \in {\cal H}}\frac{ m_j}{{||{\cal S}||}} \ge  \sum_{j=1}^{||{\cal S}||} \frac{ m_j}{{||{\cal S}||}} $  (either ${\cal H}$ includes the $||{\cal S}||$ lightest states , or due to a lack of linear independence or vanishing coefficients,   it includes some more massive states).  Moreover, it has already been shown that  $ \overleftrightarrow{\Pi}^{\cal S}(r)  \rightarrow \overleftrightarrow{\Pi}_{\rm trunc}^{\cal S}(r)$.  Together with Eq.~(\ref{last}) this implies Eq.~(\ref{finalgoal}) and thus establishes  the lemma.

\end{document}